\documentclass[conference]{IEEEtran}
\IEEEoverridecommandlockouts
\usepackage{cite}
\usepackage{amsmath,amssymb,amsfonts}
\usepackage{algorithmic}
\usepackage{graphicx}
\usepackage{textcomp}
\usepackage{xcolor}
\def\BibTeX{{\rm B\kern-.05em{\sc i\kern-.025em b}\kern-.08em
    T\kern-.1667em\lower.7ex\hbox{E}\kern-.125emX}}

\usepackage{tikz}
\usepackage{amsmath}
\usepackage{graphicx}
\usepackage{booktabs}  
\usepackage{array} 
\usepackage{enumitem}
\usepackage{multirow}
\usepackage{makecell}
\usepackage{microtype}
\usepackage{url}
\microtypecontext{spacing=nonfrench}
\usepackage[caption=false,font=footnotesize]{subfig} %
\newcommand{\RN}[1]{\textup{\uppercase\expandafter{\romannumeral#1}}}

\def\fuzzername{\textsf{AFL-ICP}}

\begin{document}

\title{AFL-ICP: Enhancing Industrial Control Protocol Reliability via Specification-Guided Fuzzing}

\author{%
\IEEEauthorblockN{%
    Jiaying Meng\textsuperscript{$\dagger$},\quad
    Xuewei Feng\textsuperscript{$\ddagger$},\quad
    Qi Li\textsuperscript{$\ddagger\dagger$},\quad
    Min Liu\textsuperscript{$\S\dagger$},\quad
    Ke Xu\textsuperscript{$\ddagger\dagger$}%
}
\IEEEauthorblockA{%
    \textsuperscript{$\dagger$}\textit{Zhongguancun Lab}\quad
    \textsuperscript{$\ddagger$}\textit{Tsinghua University}\quad
    \textsuperscript{$\S$}\textit{Institute of Computing Technology, Chinese Academy of Sciences}%
}
}

\maketitle

\begin{abstract}
Industrial Control Protocols (ICPs) are critical to the reliability and stability of industrial infrastructure, yet their security is fundamentally compromised by a \textit{specification-blindness} bottleneck. 
Modern fuzzers, constrained by observation-driven inference struggle to penetrate deep protocol states or detect subtle semantic deviations. 
In this paper, we present \fuzzername{}, an autonomous fuzzing framework that pioneers a specification-driven paradigm. \fuzzername{} features a context-aware \textit{specification formalization} pipeline to transform complex specifications into rigorous machine-executable grammars. 
Building on this formalized specification, \fuzzername{} leverages LLMs to enable automated protocol adaptation and seed
generation, allowing for rapid extension to new protocols with minimal manual effort. 
Additionally, it includes an LLM-powered differential checker that cross-references implementation outputs with specification requirements to detect subtle semantic and logic bugs that existing fuzzers cannot detect. 
We implement \fuzzername{} and evaluate it on four widely used ICPs, including both open-source and closed-source variants.
Results show that \fuzzername{} significantly outperforms state-of-the-art fuzzers in coverage and uncovers 24 previously unknown vulnerabilities, for which we have received acknowledgments from affected vendors (e.g., FreyrSCADA). Specifically, the identified vulnerabilities include 16 semantic and logic bugs that can silently disrupt industrial operations and degrade service availability.

\end{abstract}


\section{Introduction}
Industrial Control Protocols (ICPs) serve as the backbone of industrial environments, orchestrating real-time communication and command execution across heterogeneous devices. 
Early ICPs operated in physically isolated environments, 
which led to the design of many protocols without 
built-in security mechanisms such as authentication and 
encryption. 
However, as the isolation and obscurity of industrial 
networks have diminished, the number of attacks 
exploiting ICPs has rapidly increased. 
These attacks not only compromise system reliability and 
availability, but also pose significant threats to 
critical infrastructure in the transportation and energy sectors \cite{anton2021global}. 

Protocol fuzzing has emerged as a critical technique for ensuring quality of service (QoS) in industrial networks, as it proactively identifies implementation vulnerabilities that could degrade system reliability~\cite{pham2020aflnet, natella2022stateafl}. 
%
%
However, fuzzing ICPs remains fundamentally limited by a \textbf{Specification-Blindness} bottleneck. 
This cognitive gap manifests in two dimensions:

\noindent \textbf{i) The Input Dimension:}
Existing fuzzers operate as black-box or grey-box observers, inferring protocol logic solely from implementation feedback or captured traffic.
Since they lack access to the ground-truth rules defined in natural language specifications, they have to rely on \textit{inductive reasoning} from observed information to construct protocol state machines. 
This approach inherently under-fits the complex logic of ICPs, failing to discover ``dark'' states and transitions that are defined in the specification but absent in observations. 
Consequently, they struggle to generate the precise, state-aware sequences required to penetrate deep protocol logic.

\noindent \textbf{ii) The Output Dimension:} 
Traditional fuzzers rely on generic fault signals (e.g., crashes or hangs) as oracles. 
However, they lack the semantic understanding to verify whether a non-crashing response conforms to the specification's ``correctness contract.'' 
This leaves them blind to \textit{semantic deviations}, which leads to subtle failures that do not crash the system but disrupt state synchronization, bypass validation logic, or leak sensitive information, directly undermining the stability and reliability of critical infrastructure.  


We argue that the protocol specification serves as the ultimate ground-truth for correctness, yet it remains inaccessible to traditional fuzzers due to its unstructured nature. 
To break the specification-blindness bottleneck, a paradigm shift is required: from \textit{observation-driven inference} to \textit{specification-guided generation and verification} for both input and output.

Recent breakthroughs in Large Language Models (LLMs) provide the missing link for this transformation. 
LLMs possess the capability to distill rigorous formal constraints (e.g., state machines, packet schemas) from ambiguous natural language descriptions. 
This positions them not merely as auxiliary tools \cite{meng2024large}, but as the fundamental semantic bridge capable of translating human-readable standards into machine-executable fuzzing policies. 


However, unlocking this potential is non-trivial. 
Our preliminary experiments reveal that off-the-shelf LLMs struggle to formalize complex ICP standards directly. 
When fed raw specifications, LLMs frequently exhibit context rot and hallucinations, failing to maintain the strict semantic fidelity required for protocol fuzzing. 
These failures highlight two critical technical barriers:
\textbf{(C1) Inaccurate Understanding of Specifications: }
ICP specifications are lengthy and multimodal, often exceeding the effective context window of LLMs. 
Naive retrieval or summarization fragments the protocol logic, leading to misinterpretations of cross-page dependencies and state transition rules.
\textbf{(C2) Ineffective Representation of Semantics: } 
LLMs struggle to zero-shot transform unstructured specifications into the rigorous, machine-readable formats required by fuzzers. 
Without domain-specific schemas to constrain the output, generated structures are often verbose, inconsistent, or omit critical details, preventing direct integration with downstream fuzzing components. 

To surmount these barriers, we introduce a context-aware \textit{specification formalization} methodology that systematically transforms complex, multimodal ICP specifications into structured, machine-readable formats. 
To address the challenge (C1), we employ content-aware segmentation, multimodal layout reconstruction, and semantic relevance denoising.  
These mechanisms effectively filter out irrelevant noise while preserving critical grammar structures, ensuring an accurate understanding of the lengthy document.  
To resolve the challenge (C2), we design a universal protocol schema comprising three complementary formats to capture accurate protocol grammar and enforce strict correctness through an adversarial consistency verification loop. 

Building on this formalized specification, we design \fuzzername{}, an autonomous fuzzing framework that systematically integrates LLMs into every phase of the workflow. 
For the input dimension, \fuzzername{} utilizes the formalized specification to enable automated protocol adaptation and seed generation, allowing for rapid extension to new protocols with minimal manual effort. 
For the output dimension, the framework includes an LLM-powered differential checker that cross-references implementation outputs with specification requirements to detect subtle \textit{semantic and logic bugs} that existing fuzzers cannot detect. 
Evaluations on four widely-used ICPs (Modbus TCP, EtherNet/IP, IEC 104, and SLMP) show that \fuzzername{} significantly outperforms state-of-the-art fuzzers 
in both state-space and code coverage, enabling the discovery of 24 previously unknown vulnerabilities, including 16 semantic and logic bugs.


\noindent \textbf{Contributions}. Our main contributions are the following:
\begin{itemize}[leftmargin=*]

\item We introduce the first end-to-end specification-driven fuzzing framework for ICP implementations, which operationalizes natural-language specifications into machine-executable artifacts that drive both input generation and conformance checking, bridging the long-standing specification–implementation gap.

\item We integrate context-aware formalization, LLM-assisted protocol adaptation, specification-guided seed synthesis, and a specification-conformance oracle into a single coherent pipeline, enabling the detection of subtle semantic and logic flaws beyond the reach of coverage-guided fuzzing.

\item We implement and evaluate \fuzzername{} on four widely deployed ICPs, outperforming state-of-the-art fuzzers and uncovering 24 new vulnerabilities, including 16 significant semantic and logic flaws. We release the source code of \fuzzername{} at \url{https://github.com/susu3/AFL-ICP}.

\end{itemize}

\section{Motivation and Challenges}
\label{sec:background}
\subsection{Motivation: Limitations of Existing ICP Fuzzing}
\begin{figure*}[th]
    \centering
    \includegraphics[width=0.9\textwidth]{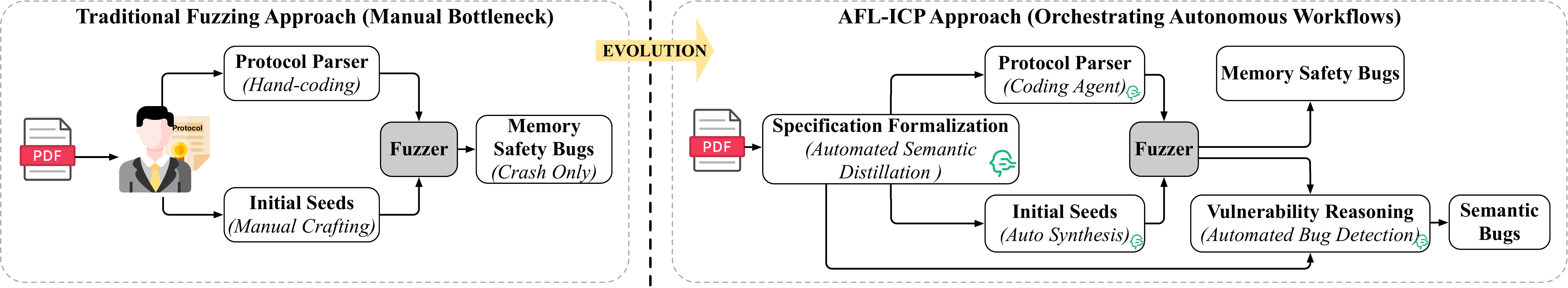}
    \vspace{-0.3cm}
    \caption{Comparison between the traditional fuzzing and the \fuzzername{} method.}
    \label{fig:motivation}
    \vspace{-0.5cm}
\end{figure*}

%
Adapting fuzzers to new ICPs mandates a labor-intensive, human-in-the-loop workflow (Fig. \ref{fig:motivation}). 
Security analysts must manually dissect voluminous specifications, hand-craft protocol drivers, and curate seed corpora. 
Only after this exhaustive preparation can the automated fuzzing loop of testcase generation, execution, and feedback collection commence. 
Even then, it remains confined to detecting shallow memory safety violations.
Our work redefines this traditional workflow by introducing an AI-native, specification-driven paradigm.
This transformation yields two distinct advantages:

\noindent\textbf{i) Orchestrating Autonomous Workflows:} 
We elevate pre-fuzzing stages from labor-intensive manual tasks to autonomous or AI-accelerated processes. 
Specifically, we transition specification parsing, protocol adaptation, and initial seed generation to purely AI-driven pipelines.
This fundamentally mitigates the scalability bottlenecks inherent in manual fuzzing campaigns. 

\noindent\textbf{ii) Expanding vulnerability Detection Capabilities:} 
While traditional fuzzers are constrained to identifying memory safety violations, our specification-aware design enables the discovery of semantic and logic bugs, directly assuring protocol QoS. 

\subsection{Challenges in ICPs Fuzzing}
\noindent\textbf{i) Specification Parsing: The Multimodal Barrier.}
Automating ICP specification processing faces unique challenges. 
First, the sheer scale exceeds the context window of current models. 
Unlike the HTTP/1.1 standard (RFC 7230--7235, 305 pages, 380 KB in PDF), ICP specifications are voluminous (e.g., EtherNet/IP spans 1,444 pages, 7.6 MB in PDF), making direct processing prohibitive and leading to context rot. 
Second, unlike standardized IETF protocols, ICP 
specifications are often distributed as unstructured PDFs with critical protocol information frequently presented through non-textual elements such as tabular and architectural diagrams.
Naive text extraction fails to preserve the layout of these elements, leading to loss of semantic information.
Furthermore, many ICPs, originating from legacy bus architectures, exhibit complex characteristics, including numerous information fields, multilayer nested structures, and fine-grained control mechanisms. 
LLMs often struggle with these intricate details and subtle requirements. 

\noindent\textbf{ii) Protocol Adaptation: High Engineering Overhead.}
Extending fuzzing to new ICPs necessitates the manual implementation of serialization and state management logic. 
This is prohibitively expensive due to the proprietary nature and diverse message formats of ICPs. 
Moreover, a major challenge lies in the limited observability of protocol states. While standard protocols provide immediate feedback through explicit status fields (e.g., HTTP response codes), ICPs typically lack such standardized state signaling. 
This forces analysts to rely on deep domain expertise to manually map message sequences to protocol states, a tedious process that severely hinders scalability.

\noindent\textbf{iii) Initial Seeds: The Diversity and Validity Deficit.}
High-quality seed corpus are crucial for maximizing fuzzing coverage \cite{klees2018evaluating, herrera2021seed}. 
However, autonomously generating such seeds via LLMs faces a dual challenge rather than a simple trade-off.
Without explicit grammar constraints, even advanced reasoning models struggle to produce structurally valid packets that strictly adhere to complex binary protocols. 
Simultaneously, models tend to converge on ``typical'' communication patterns, failing to generate the diverse edge cases required to penetrate deep program states. 
Consequently, direct LLM generation fails to produce a seed corpus that is both structurally valid and sufficiently diverse to penetrate deep program states. 

\noindent\textbf{iv) Vulnerability Reasoning: The Semantic Blind Spot.}
Current fuzzers rely on generic sanitizers \cite{ubsan2014,serebryany2012addresssanitizer,song2019sok} 
that are effective for memory safety but blind to semantic deviations. 
Logic bugs, such as incorrect state transitions or silent packet drops, do not trigger crashes but severely compromise protocol reliability. 
Manually encoding these complex rules into checkers creates a formalization bottleneck, as translating ambiguous specification text into precise verification logic is error-prone.

\section{Design and Methodology}
\subsection{Overview}
\begin{figure*}[t]
    \centering
    \includegraphics[width=0.9\textwidth]{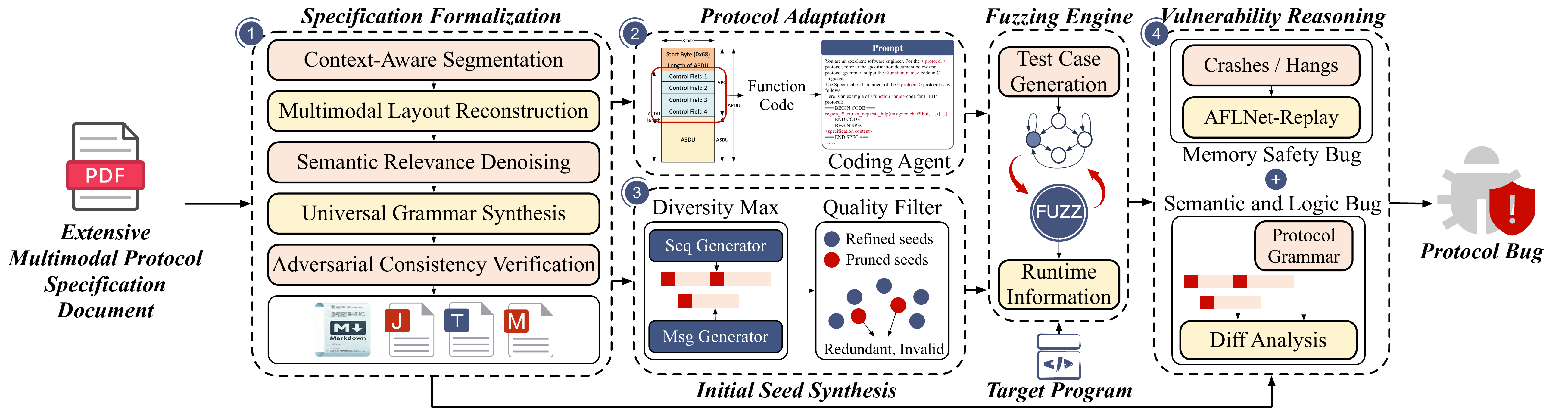}
    \caption{Architecture of \fuzzername{}: end-to-end workflow from multimodal specification formalization to LLM-assisted vulnerability reasoning.}
    \label{fig:model}
    \vspace{-0.5cm}
\end{figure*}

To bridge the gap between abstract specifications and concrete implementations, we present \fuzzername{}, an AI-native fuzzing architecture that transforms the traditional human-dependent workflow into an autonomous, specification-driven pipeline.
Fig. \ref{fig:model} illustrates the system architecture.
By systematically integrating LLMs, \fuzzername{} automates the labor-intensive tasks of specification analysis and seed generation, extending the fault detection horizon beyond simple crashes to complex semantic and logic bugs.
The system operates through three cohesive phases:
\begin{enumerate}[label=\textbf{Phase \Roman*:}, wide=0pt]
    \item \textbf{Autonomous Offline Preparation.} \fuzzername{} initiates with a \textit{Specification Parsing} pipeline (Sec. \ref{sec:parse}) that extracts structured grammar from multimodal PDF specifications. Leveraging this grammar, the \textit{Protocol Adaptation} module (Sec. \ref{sec:code}) synthesizes protocol-specific parsing logic, while the \textit{Initial Seed Generation} engine (Sec. \ref{sec:seed}) constructs a diverse, valid corpus to bootstrap the campaign—effectively eliminating the ``cold start'' bottleneck of manual setups.
    \item \textbf{Online Fuzzing.} The mutation engine executes iterative exploration of the target state space. Distinct from traditional fuzzers, it continuously logs deep semantic interactions and potential state deviations alongside standard crash signals, creating a rich behavioral dataset for subsequent analysis.
    \item \textbf{Memory and Semantic Bug Reasoning.} In the final phase, our analysis engine performs a dual-layer evaluation: it triages standard memory-safety violations and deploys the \textit{Semantic Bug Reasoning} oracle (Sec.~\ref{sec:bug}) to uncover subtle logical non-conformances by cross-referencing execution traces against the formal specification.
\end{enumerate}

The remainder of this section elaborates on the core components of the \fuzzername{} architecture.

\subsection{Specification Formalization}\label{sec:parse}
Automating the interpretation of ICPs is not merely a parsing task but a \textit{domain formalization} challenge: transforming high-entropy, ambiguous natural language into low-entropy, executable protocol grammar.
This transformation faces three entropic barriers:
\textit{(1) Information Sparsity:} The normative constraints (message formats, state machines) are often buried within massive volumes of irrelevant descriptions (e.g., security policies and theoretical background).
\textit{(2) Multimodal Fragmentation:} Critical logic is often distributed across textual definitions and visual artifacts (tables, diagrams), necessitating semantic integration.
\textit{(3) Ambiguity vs. Precision:} Natural language specifications rely on implicit context, whereas fuzzers require explicit, machine-readable constraints.

To bridge this gap, we propose a five-stage \textit{automated specification formalization} pipeline. 
This pipeline functions as a knowledge distillation engine, progressively refining unstructured PDF data into a rigorous \textit{universal protocol schema}. 
Crucially, this process also yields \textit{filtered relevant documentation}, a denoised, logic-dense intermediate artifact.
Both the structured schema and this refined textual context serve as essential inputs for subsequent seeding, adaptation, and reasoning tasks. 
The following paragraphs detail the formalization stages. 

\textbf{Stage 1: Context-Aware Segmentation.}
To accommodate LLM context window constraints without sacrificing \textit{local semantic coherence}, we perform logic-aware segmentation. 
Instead of arbitrary token chunking, which often severs critical cross-page references, we utilize ``structural anchors'' (e.g., Volumes, Chapters, Sections) to partition the voluminous specification. 
We strictly align segment boundaries with natural semantic breaks, ensuring that contextually dense units, such as large message definition tables spanning multiple pages,remain within a single processing block. 
For exceptionally large sections ( $>50$ pages), we implement secondary partitioning at lower-level headings to maintain processability.

\textbf{Stage 2: Multimodal Layout Reconstruction.}
Standard OCR destroys the spatial semantics essential for understanding ICPs (e.g., bit-map diagrams). 
We resolve this incompatibility through a dual-stream \textit{visual-textual fusion} pipeline. 
Our methodology leverages Mistral OCR~\cite{mistral2025ocr} to recover the textual layer, converting PDF documents into Markdown format. 
This process preserves the original structure, including chapter organization, text, and tables, while also correctly placing images. 
Critical visual elements such as architectural diagrams and protocol illustrations are then processed through vision LLMs like GPT-4o\cite{openai2024gpt4o}, based on the insight that visual constraints are indispensable for reconstructing message formats and state machine topology.
 
\begin{figure}[htbp]
    \vspace{-0.3cm}
    \centering
    \includegraphics[width=0.48\textwidth]{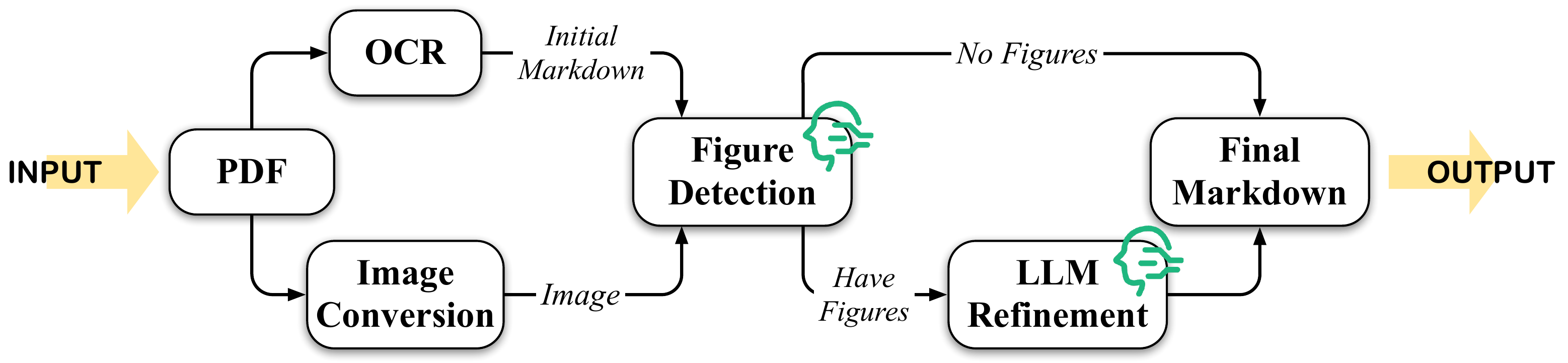}
    \vspace{-0.1cm}
    \caption{The Multimodal Layout Reconstruction Pipeline}
    \vspace{-0.3cm}
    \label{fig:pdftomd}
\end{figure}

Fig. \ref{fig:pdftomd} illustrates this multimodal document conversion pipeline: 
Initial OCR conversion generates a baseline Markdown representation for each PDF page. 
Upon figure detection (identified through pattern matching, e.g., ``Figure x''), we feed both the initial Markdown (providing essential textual context) and the corresponding image extracted from the PDF (via PyMuPDF) to a vision LLM. 
The vision LLM then generates a descriptive text representation of the visual logic (e.g., state diagrams and packet formats). 
The resulting textual descriptions are subsequently merged into the initial Markdown, producing a unified, comprehensive document suitable for subsequent LLM processing. 
The resulting Markdown is then input into a text-based LLM for further processing steps.

\textbf{Stage 3: Semantic Relevance Denoising.} 
We treat relevance filtering as a signal-to-noise problem: filtering out \textit{auxiliary noise} (general introductions, physical layer specs, deployment guides) to amplify \textit{core signal} (frame structures, data encoding, and state logic). 
Despite their comprehensive nature, ICP specifications contain substantial content not directly pertinent to protocol formatting and operational rules. 
To replace the traditional, time-consuming, and error-prone manual screening process, we leverage LLMs to perform this semantic filtering. 

We observed that chapters relevant to protocol grammar often exhibit consistent terminological patterns, specifically including keywords such as \textit{encapsulation}, \textit{messages}, \textit{data flow}, \textit{layering}, \textit{frame}, and \textit{format}. 
We leverage this observation by embedding these terminological markers into carefully constructed prompts, enabling the LLM to systematically identify and extract pertinent sections with high precision. 
The content filtered in this stage establishes the foundational corpus for all subsequent processing phases.

\textbf{Stage 4: Universal Grammar Synthesis.} 
This stage employs LLMs to systematically extract structured protocol grammar that serves as a reference and guide for subsequent seed generation and vulnerability reasoning. 
However, this process faces two key challenges: \textit{what specific information to extract} and \textit{how to design a universal storage format} that can accommodate diverse ICPs while remaining machine-readable. 

\textbf{Challenge 1: Information Selection (The Client-Side Perspective).} 
Protocol fuzzing operates as a client-side activity, sending crafted requests to trigger server vulnerabilities. 
This requires a specific subset of knowledge focused on \textit{active message construction} rather than passive analysis. 
Therefore, we restrict our extraction to information critical for simulating a valid client: precise field layouts for message serialization, dependency logic for valid payload construction, and state transition rules for navigating protocol sessions. 

\textbf{Challenge 2: Universal Representation.} 
ICPs exhibit vast structural diversity (e.g., Modbus has simple frames, while IEC104 has complex nested types). 
To map these heterogeneous inputs into a agent-readable model, we synthesize a \textit{universal protocol schema} composed of three orthogonal representations:

\textbf{1) Packet Structure Specification (JSON Schema):} To capture static message definitions, we utilize JSON schema. 
This format allows us to enforce rigid syntactic rules, detailing \textit{field sequences, data types (integer, string), endianness conventions, valid value ranges, and the hierarchical organization of headers}. 
This provides the explicit constraint structure required for accurate bit-level message construction and mutation. 

\textbf{2) Protocol State Machine (Mermaid Notation):} 
To capture temporal logic, we translate \textit{client-side state transition rules} into Mermaid state diagrams. 
This text-based diagramming language allows us to represent state definitions, transition conditions, and event triggers (including state-dependent message variations) in a format that is both human-readable and topologically parsable, guiding the generation of realistic multi-step communication flows. 

\textbf{3) Contextual Dependencies (Structured Markdown):} Complex ``fuzzy'' logic, such as \textit{conditional field presence based on flag values, cross-field validation rules, and implicit behavioral requirements}, often defies rigid schema definition. 
We capture these nuances by summarizing them into a structured Markdown document. 
This retains the semantic richness of the original natural language constraints while providing a clean, retrieval-ready context for the coding agent to handle edge cases. 

To operationalize this multi-format storage strategy, we construct structured prompts that integrate the predefined storage schemas with the filtered protocol documentation from Stage 3, guiding the LLM to systematically parse the protocol specifications and populate the corresponding data structures. 
The prompt design incorporates explicit formatting requirements for each storage type, enabling the LLM to transform unstructured protocol documentation into structured data. 
The extracted and standardized grammar representations can be directly utilized as a reminder or reference in subsequent automated processing stages.

\textbf{Stage 5: Adversarial Consistency Verification.} 
LLMs are prone to hallucinations or omissions.  
To mitigate this, we implement a \textit{cross-representation verification loop} that prompts the LLM to cross-reference the extracted universal grammar against the filtered source documentation rather than relying on unaided self-reflection.
This stage implements a systematic verification process that performs secondary validation of the protocol grammar extracted in Stage 4. 
We construct structured validation prompts that embed the generated text descriptions, JSON schemas, and Mermaid diagrams alongside the filtered protocol documentation from Stage 3 (in Markdown format), instructing the LLM to perform cross-referential consistency checking. 

Specifically, for JSON Schemas, the LLM scrutinizes syntactic and semantic correctness, identifying discrepancies in field definitions, data types, and dependencies against the source documentation. 
Similarly, for Mermaid state machines, it verifies topological consistency by cross-referencing state transitions and event triggers. 
When inconsistencies are detected, the LLM generates detailed diagnostic reports specifying the exact field locations, nature of errors, and suggested corrections. 


Upon error detection, we re-invoke the LLM with the diagnostic feedback to regenerate corrected representations.
This adversarial loop enables systematic error detection and correction across all storage formats, demonstrating measurable improvements in extraction accuracy.

\subsection{Protocol Adaptation}\label{sec:code}
Adapting stateful fuzzers to binary ICPs necessitates bridging the gap between raw byte streams and structured protocol semantics. 
For instance, adapting an AFLNet-style stateful fuzzer requires the agent to generate two protocol-specific functions: one that delimits message boundaries within the request byte stream, and one that extracts the state identifier from server responses. 
To automate this, we introduce a fully automated protocol adaptation methodology driven by a multi-agent system. This pipeline coordinates a \textit{coding agent} to synthesize parsing logic, with \textit{code review} and \textit{integration test agents} that iteratively validate and refine the generated code, ensuring a robust and autonomous adaptation process.

\textbf{Function Code as State Identifier.}
A key challenge in adapting fuzzers to ICPs lies in state management, as most fuzzing tools rely on state codes to track protocol execution states and guide mutation strategies.
ICPs are characterized by an \textit{operation-centric} design philosophy: their state transitions are encoded directly in the operation type of each message---typically expressed as function codes or command codes---rather than maintained through implicit context.
We identify a critical insight: \textit{In ICPs, function codes are not merely operation indicators but the primary carriers of protocol state---session lifecycle (e.g., RegisterSession / UnRegisterSession) and data-transfer activation (e.g., STARTDT / STOPDT) are both realized through function-code transitions.}
Therefore, we implement a mapping strategy that uses function codes as state identifiers, enabling the fuzzer to maintain protocol state awareness.
For protocols with composite operation fields (e.g., SLMP's command and sub-command pair), we directly sum the two fields to derive a unified state identifier.
This design naturally aligns with the inherent semantics of ICPs while maintaining compatibility with existing state-based fuzzing mechanisms.

\textbf{Coding Agent.}
Our approach employs a SOTA coding agent (e.g., Claude Code) to generate protocol support code via a structured prompt engineering methodology. 
We construct comprehensive prompts that integrate three essential components: (1) existing protocol support code templates that demonstrate the required code structure and API interfaces, (2) the extracted protocol grammar from Stage 4, including JSON schemas, state machine descriptions, and field dependencies, and (3) the filtered protocol documentation from Stage 3 to provide additional contextual information.
The coding agent receives these inputs and generates protocol-specific parsing code that conforms to the established code patterns while implementing the target protocol's unique characteristics.

\textbf{Automated Verification and Refinement.}
To ensure the reliability of the generated code without human intervention, we deploy two specialized agents.
First, the \textit{code review agent} performs static analysis on the generated code. 
It checks for compliance with coding standards, potential logic errors, and security vulnerabilities (e.g., buffer overflows in the parsing logic itself).
Second, the \textit{integration test agent} dynamically validates the code. 
It attempts to compile the new ICP adapter and link it with the core fuzzing engine. 
Upon successful compilation, it runs the fuzzer against a small set of valid seeds to verify basic functional correctness (e.g., successfully parsing a valid packet without crashing). 
If either agent detects an issue, detailed feedback is automatically fed back to the \textit{coding agent}. The coding agent then analyzes the feedback, revises the code, and resubmits it for verification. 
This iterative self-correction loop continues until the protocol adapter passes all checks, achieving a fully autonomous generation pipeline.

\subsection{Initial Seed Synthesis}\label{sec:seed}
It is widely recognized that initial seed corpus quality significantly impacts fuzzing's effectiveness \cite{klees2018evaluating, herrera2021seed}. 
High-quality seeds encompass diverse input formats of the tested protocol, enabling the fuzzing tools to rapidly explore deeper and broader code paths, thus improving overall test coverage. 
Moreover, carefully selected seeds reduce the learning overhead for fuzzers, facilitate the efficient construction of protocol state machines, and help generate valid, protocol-conformant mutated packets. 
Motivated by the generative capabilities of LLMs, we propose an automated pipeline to synthesize the seed corpus.

Generating effective seeds via LLMs faces a dual deficit in \textit{validity} and \textit{diversity}. 
Without explicit constraints, models struggle to produce structurally valid packets for complex protocols, while simultaneously converging on repetitive patterns that miss critical edge cases.

To address this challenge, we implement a two-stage seed generation methodology that decouples diversity maximization from quality assurance. 
The first stage focuses on \textit{diversity maximization}.
We perform multiple queries to generate a broad spectrum of seed candidates. 
To guide this generation, our prompts synthesize two primary information sources: (1) the extracted universal protocol schema (grammar constraints), and (2) the filtered relevant documentation (semantic context). 
Additionally, if pre-captured network traffic is available, it can be included as few-shot examples.
This combination allows the model to produce seeds that mirror authentic communication patterns while maintaining the randomness for fuzzing.

The second stage acts as a \textit{quality filter}.
We utilize standard seed optimization tools (e.g., \texttt{afl-tmin}) to prune the raw generated corpus. 
This step eliminates redundant, invalid, or oversized seeds, retaining only a compact set of high-quality inputs. 
By effectively separating the ``creative'' generation phase from the ``restrictive'' validation phase, we achieve a final seed corpus that simultaneously ensures high structural diversity and strict protocol compliance.

Within this two-stage framework, we synthesize seeds at two distinct granularities: message-level seeds and sequence-level seeds. 
Message-level seeds consist of individual protocol packets that test specific protocol features and edge cases. 
Sequence-level seeds comprise multi-packet communication flows that exercise protocol state transitions and complex interaction patterns. 
To facilitate automated extraction, we enforce structured XML-like output formats via prompt constraints: message-level seeds are enclosed in \texttt{<sequence>} tags (treated as single-step sequences), while sequence-level seeds use nested structures as follows: \texttt{<sequence><message>...}\allowbreak\texttt{</message>...</sequence>}. 

For sequence generation, our approach explicitly instructs the LLM to first traverse a valid execution path on the protocol state machine, then systematically populate each packet in the identified path with appropriate field values.
This structured generation ensures that the resulting sequences strictly adhere to the state machine topology, guaranteeing logical continuity between consecutive messages.

\subsection{Semantic Vulnerability Reasoning}\label{sec:bug}
While existing sanitizers effectively capture memory safety violations, they remain blind to semantic vulnerabilities where implementations deviate from protocol rules without crashing. 
 
To detect these deep logic bugs, we introduce \textit{specification-implementation differential analysis}.
Unlike traditional differential testing, which compares the outputs of two software binaries, our approach treats the \textit{protocol specification itself} as the absolute ``golden standard''. 
We propose a \textit{specification-conformance oracle} that leverages LLMs to directly cross-reference the fuzzer's interaction history (both request and response packets) against the formal requirements defined in the specification.

We employ a dual-layer vulnerability detection strategy: utilizing traditional sanitizers for memory safety bugs while deploying our specification-conformance oracle for semantic and logic bugs. 
When the implementation's behavior contradicts the formal definition, the oracle flags the discrepancy.
To ensure comprehensive coverage, our oracle operates on two distinct recording mechanisms:

\textbf{Path-Triggered Recording.} When a new execution path is discovered, we capture the complete communication history. 
This allows us to verify conformance for every unique functional state reached by the fuzzer, ensuring that new logic paths adhere to specification constraints.

\textbf{Probabilistic Sampling.} Logic bugs often manifest in paths already explored (e.g., incorrect error codes) without triggering new coverage. 
To capture these, we employ a probabilistic sampling strategy that periodically records test cases regardless of coverage gain, ensuring that ``silent'' logic violations do not escape detection.

Invoking LLMs for real-time bug analysis during fuzzing would significantly slow down the testing process, we adopt a record-then-check approach to maintain fuzzing efficiency.
During the fuzzing phase, we only record test cases and their execution traces; after fuzzing terminates, we invoke the LLM-enabled specification-conformance oracle to analyze all recorded test cases in batch.
This deferred analysis strategy ensures that fuzzing throughput remains unaffected while still leveraging the power of LLMs for semantic bug detection.

Concretely, each recorded test case---comprising the request packet and the corresponding server response---is provided to the LLM together with the relevant specification context (the filtered Markdown documentation from Stage 3 and the structured grammar from Stage 4). The prompt instructs the LLM to flag any behavioral deviation and to ground each flagged finding in a specific specification clause; requiring this explicit clause citation discourages ungrounded hallucinations. Any residual false positives are caught by the manual validation pipeline described in Sec.~\ref{bug2}.

\section{Implementation}
We have implemented a prototype of \fuzzername{} based on AFLNet, a state-of-the-art state-guided greybox fuzzer for network protocols. 
Our implementation consists of over 6,000 lines of code. The core fuzzing components are written in C and integrated directly into AFLNet, while the 
Stage 1-3 of specification formalization modules are implemented in Python. 

\textbf{LLM Integration.}
We strategically employ different LLMs based on their architectural strengths.
For the \textit{multimodal layout reconstruction} task (Sec.~\ref{sec:parse} Stage 2), we utilize \textit{GPT-4o} due to its superior vision-language capabilities in interpreting complex diagrams.
For \textit{protocol adaptation} (Sec.~\ref{sec:code}), the coding agent is instantiated with \textit{Claude Code}, which provides agentic code generation, review, and execution feedback in a single loop.
For all other text-centric tasks, including semantic relevance denoising, universal grammar synthesis, seed synthesis, and bug reasoning, we employ \textit{Gemini-2.5-Pro}, leveraging its extensive context window and reasoning performance.
All interactions are managed through their respective native APIs with robust error handling and retry mechanisms.

The pipeline is intentionally model-agnostic: each stage prescribes a capability requirement  rather than a specific vendor. The proprietary models above were chosen for their leading capability at the time of evaluation, but any model that meets the corresponding capability bar can be substituted in principle. 
We further note that LLM invocations are confined to the offline preparation and post-campaign analysis stages; the online fuzzing loop itself involves no LLM calls.

\textbf{Document Processing.}
Our specification formalization pipeline integrates multiple tools to handle complex PDF documents.
We employ Mistral OCR for converting PDF pages into Markdown format. 
We utilize PyMuPDF (also known as fitz) to extract embedded artifacts for the vision LLM pipeline. 
These components work synergistically to transform unstructured PDF specifications into final Markdown. 

\textbf{Fuzzing Infrastructure Enhancement.}
To facilitate the two-stage seed generation (Sec.~\ref{sec:seed}), we integrated \texttt{afl-tmin} into our pipeline for automated corpus quality filter.
For semantic bug reasoning (Sec.~\ref{sec:bug}), we implemented a lightweight trace logger within AFLNet's main loop that records request-response pairs to disk for post-campaign batch analysis, so that no LLM invocation occurs inside the fuzzing loop.

\textbf{Protocol Support.}
We have extended AFLNet to support four widely-used ICPs: Modbus TCP, EtherNet/IP, IEC 104, and SLMP.
For each protocol, our \textit{coding agent} successfully synthesized protocol-specific message parsers, state machine handlers, and response validators.
The generated support code integrates seamlessly with AFLNet's existing infrastructure, enabling state-guided mutation and coverage tracking for these binary ICPs.

\section{Evaluation}
To evaluate the effectiveness of \fuzzername{}, we try to answer the following questions:

\textbf{Q1. State Coverage.} 
Does \fuzzername{} explore more protocol states compared to baselines?

\textbf{Q2. Code Coverage.} 
How much more code coverage does \fuzzername{} achieve compared to the baseline? 

\textbf{Q3. Ablation Study.} What is the impact of the each component on the performance of \fuzzername{}? 

\textbf{Q4. Bug Identification.} 
Can \fuzzername{} detect previously unknown memory safety and semantic and logic bugs?

\subsection{Experimental Design}
\textbf{Target Programs.} 
To evaluate \fuzzername{}, we conducted experiments on seven mature implementations across four widely used ICPs: \textit{libmodbus} and \textit{libplctag} for Modbus TCP; \textit{OpENer} and \textit{EIPScanner} for EtherNet/IP; \textit{FreyrSCADA} and \textit{IEC104} for IEC 104; and \textit{libslmp2} for SLMP. 
These implementations are widely used 
both in enterprises and individual users. 
Some implementations can be directly used as test 
subjects without modification, while others require 
additional components to construct a testable 
environment. 
For example, EIPScanner\cite{eipscanner} is a 
client-side library and cannot be evaluated 
independently; therefore, we implemented a 
server-side program that reuses its protocol parsing 
and handling logic, enabling effective evaluation 
within our testing framework. 

\textbf{Baselines.}
We compare \fuzzername{} against two state-of-the-art open-source fuzzers: \textit{AFLNet}~\cite{pham2020aflnet} (a mutation-based, state-guided fuzzer) and \textit{ChatAFL}~\cite{meng2024large} (an LLM-guided fuzzer).
Since neither tool natively supports ICPs, we extend both with the same protocol adapter code generated by our coding agent, ensuring all three fuzzers operate on identical protocol parsing logic so that any coverage difference reflects the fuzzing strategy rather than harness quality.
Both baselines were configured with optimal parameters to ensure a fair comparison.
We note that prior ICP-specific fuzzers such as Polar~\cite{luo2019polar} and related works~\cite{luo2020ics, zuo2022vulnerability} are not included as baselines because their source code is not publicly available, precluding direct empirical comparison.



\textbf{Evaluation Methodology.} 
We evaluate effectiveness based on both coverage and vulnerability discovery. 
For coverage, we report the coverage of both the code and the state space. 
Code coverage (branch and line) is measured using \textit{gcovr} \cite{gcovr}, with the exception of FreyrSCADA \cite{iec104-2} whose core logic resides in a closed-source precompiled library \texttt{libx86\_x64-iec014.a}. 
State-space coverage is quantified by (1) state coverage, representing the number of distinct protocol states explored, and (2) transition coverage, reflecting the diversity of state transition paths exercised. 
Both are extracted from the fuzzer's \texttt{n\_nodes} and \texttt{n\_edges} outputs.
To ensure statistical significance and mitigate the impact of non-determinism, all results are averaged over five independent 24-hour repetitions, with outliers excluded from the final calculation. 
Memory safety vulnerabilities are identified using ASAN \cite{serebryany2012addresssanitizer}. 
Crash-inducing sequences are reproduced via \textit{AFLNet-replay} for root-cause analysis, and unique bugs are distinguished through stack trace analysis. 
Furthermore, we identify semantic and logic vulnerabilities by utilizing our reasoning engine to flag behavioral deviations from protocol specifications for subsequent verification.

\textbf{Experimental Environment.} 
All experiments were conducted on a server running Ubuntu 20.04.6 LTS, equipped with a 28-core CPU and 32GB of RAM.

\subsection{Experimental Results}
\begin{table*}[thbp]
    \centering
    \caption{Comparison of average state and transition coverage between \fuzzername{} and baselines.}
    \vspace{-0.2cm}
    \resizebox{\textwidth}{!}{
    \begin{tabular}{l|cc |cccc |cccc }
        \toprule
        \multirow{2}{*}{\textbf{Program}} & \multicolumn{2}{c|}{\textbf{\fuzzername{}}} & \multicolumn{4}{c|}{\textbf{AFLNet}} & \multicolumn{4}{c}{\textbf{ChatAFL}} \\
        \cline{2-11}
        &{State} & {Transition}& {State} & {Improve} & {Transition} & {Improve} & {State} & {Improve} & {Transition} & {Improve} \\
        \hline
        \hline
        libmodbus &\textbf{23.60}  &\textbf{220.40}  &20.60  &14.56\%  &154.80  &42.38\%  &22.60 &4.42\% &209.20 &5.35\%\\
        libplctag &\textbf{46.20}  &\textbf{422.40}  &39.00  &18.46\%  &371.20  &13.79\%  &42.40 &8.96\% &411.60 &2.62\%\\
        OpENer &\textbf{65.40}  &\textbf{276.60}  &54.20  &20.66\%  &236.20  &17.10\%  &33.80 &93.49\%  &153.00 &80.78\%\\
        EIPScanner &\textbf{13.20}  &\textbf{14.40}  &9.00  &46.67\%  &8.00  &80.00\% &9.20 &43.48\% &8.20 &75.61\%\\
        FreyrSCADA &\textbf{14.80}  &\textbf{23.40}  &5.00  &196.00\%  &6.40  &265.63\% &5.00 &196.00\% &6.60 &254.55\% \\
        IEC104 &\textbf{7.60}  &\textbf{6.60}  &6.00  &26.67\%  &5.00 &32.00\% &6.00 &26.67\% &5.00 &32.00\%\\
        libslmp2 &\textbf{1.00}  &\textbf{1.00}  &\textbf{1.00}  &0.00\%  &\textbf{1.00}  &0.00\% &\textbf{1.00} &0.00\% &\textbf{1.00} &0.00\%\\
        \hline
        \hline
        AVG &  &  &  &46.15\%  &  &64.41\% & &53.29\% & &64.42\%\\
        \bottomrule
    \end{tabular}
    }
    \label{tab:state}
\end{table*}
\begin{table*}[thbp]
    \vspace{-0.2cm}
    \centering
    \caption{Comparison of average branch and line coverage between \fuzzername{} and baselines.}
    \vspace{-0.2cm}
    \resizebox{\textwidth}{!}{
    \begin{tabular}{l|cc |cccc |cccc }
        \toprule
        \multirow{2}{*}{\textbf{Program}} &\multicolumn{2}{c|}{\textbf{\fuzzername{}}} &\multicolumn{4}{c|}{\textbf{AFLNet}} & \multicolumn{4}{c}{\textbf{ChatAFL}} \\
        \cline{2-11}
        &{Line} &{Branch} &{Line} & {Improve} &{Branch} & {Improve} &{Line} & {Improve} &{Branch} & {Improve}\\
        \hline
        \hline
        libmodbus &\textbf{477.60}  &\textbf{216.00}  &457.60  &4.37\%  &198.40  &8.87\%  &475.00  &0.55\%  &215.60  &0.19\% \\
        libplctag &\textbf{671.00}  &\textbf{307.00}  &\textbf{671.00}  &0.00\%  &\textbf{307.00}  &0.00\%  &\textbf{671.00}  &0.00\%  &\textbf{307.00}  &0.00\% \\
        OpENer &\textbf{1412.00}  &\textbf{371.00}  &1175.00  &20.17\%  &301.60  &23.01\% &1179.80 &19.68\% &303.60 &22.20\%\\
        EIPScanner &\textbf{198.00}  &\textbf{137.00}  &180.00  &10.00\%  &121.80  &12.48\% &177.60 &11.49\% &117.40 &16.70\%\\
        IEC104 &\textbf{351.40} &\textbf{80.00}  &254.80  &37.91\%   &61.40  &30.29\%   &250.20 &40.45\% &61.60 &29.87\% \\ 
        libslmp2 &\textbf{683.25}  &\textbf{303.00}  &659.00  &3.68\%  &283.00  &7.07\% &670.00 &1.98\% &291.80 &3.84\% \\
        \hline
        \hline
        AVG &  &  &  &12.69\%  &  &13.62\% & &12.36\% & &12.13\%\\
        \bottomrule
    \end{tabular}
    }
    \vspace{-0.3cm}
    \label{tab:code}
\end{table*}
\subsubsection{State-space Coverage}\label{state}
Table \ref{tab:state} details the state and transition coverage. 
Overall, \fuzzername{} consistently outperforms baselines, achieving average improvements of 46.15\%/53.29\% in state coverage and 64.41\%/64.42\% in transition coverage compared to AFLNet and ChatAFL, respectively. 

\textbf{State Coverage.} 
\fuzzername{} excels in deep state exploration, particularly in FreyrSCADA, where it achieves a 196\% improvement. 
FreyrSCADA is closed-source and uninstrumentable; traditional fuzzers fail to efficiently guide the generation of valid sequences. 
\fuzzername{} overcomes this by initializing campaigns with offline-synthesized, specification-compliant seeds that directly penetrate deep states. 
In contrast, ChatAFL occasionally degrades (e.g., in OpENer) as general LLMs lack detailed knowledge of ICPs and thus misguide state exploration.
Note that absolute state counts vary by protocol complexity and implementation details, yet even for the extensively fuzzed libmodbus, \fuzzername{} still uncovers 14.56\% more states, confirming its ability to reach deep-seated logic. 

\textbf{Transition Coverage.} 
\fuzzername{} demonstrates even greater advantages in transition coverage, achieving a 265.63\% improvement in FreyrSCADA. 
This is primarily because \fuzzername{} overcomes the ``initial handshake barrier'': by seeding the fuzzer with valid handshake sequences derived from specifications, it immediately reaches core protocol logic, whereas traditional fuzzers waste hours mutating packets just to pass the first validation check. 
Consequently, it exercises deep logic paths that remain unreachable to blind mutation. 
Note that simple ICPs like libslmp2 show a saturation effect, where limited logical density leaves no room for further optimization. 

\subsubsection{Code Coverage.}\label{code}
Table \ref{tab:code} details the branch and line coverage.  
Overall, \fuzzername{} consistently outperforms baselines, achieving average improvements of 12.69\%/12.36\% in line coverage and 13.62\%/12.13\% in branch coverage compared to AFLNet and ChatAFL, respectively. 
Note that coverage for some targets (e.g., libplctag) hits a plateau because the provided test harnesses only expose limited functional interfaces, leaving large portions of the library unreachable regardless of fuzzer efficiency. 

\textbf{Line Coverage.} 
\fuzzername{} shows significant gains where harnesses provide broader reach, such as in OpENer and IEC104, where it achieves improvements of 20.17\% and 37.91\% over AFLNet, respectively. 
Notably, the marginal performance degradation of ChatAFL compared to the baseline AFLNet in EIPScanner and IEC104 likely stems from the stochastic nature of greybox fuzzing rather than an algorithmic deficiency, as the differences in covered lines are negligible. 

\textbf{Branch Coverage.} 
\fuzzername{} reaches deeper logic by using valid seeds to bypass gated checks. 
Notably, a ``State-Code Paradox'' appears in OpENer: ChatAFL achieves slightly higher branch coverage than AFLNet despite significantly lower state coverage. 
This indicates that ChatAFL's mutations trigger shallow error-handling branches but fail to maintain the complex session states required for deep protocol exploration. 
\fuzzername{} effectively overcomes this limitation.  


\subsubsection{Ablation Study}\label{ablation}
\begin{figure*}[t]
    \centering
    \subfloat[State Coverage\label{fig:ablation_state}]{
        \includegraphics[width=0.318\textwidth]{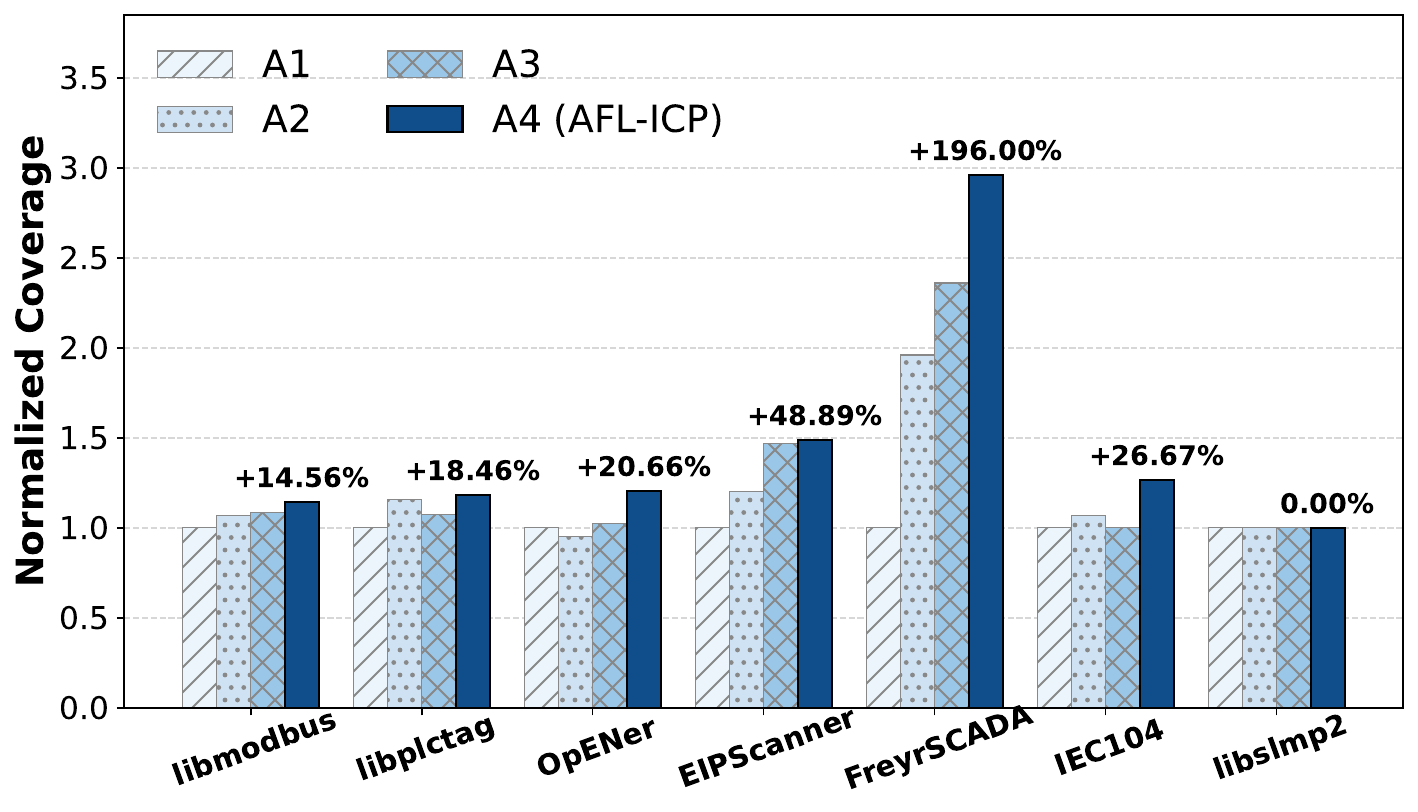}
    }
    \hfill
    \subfloat[Transition Coverage\label{fig:ablation_trans}]{
        \includegraphics[width=0.318\textwidth]{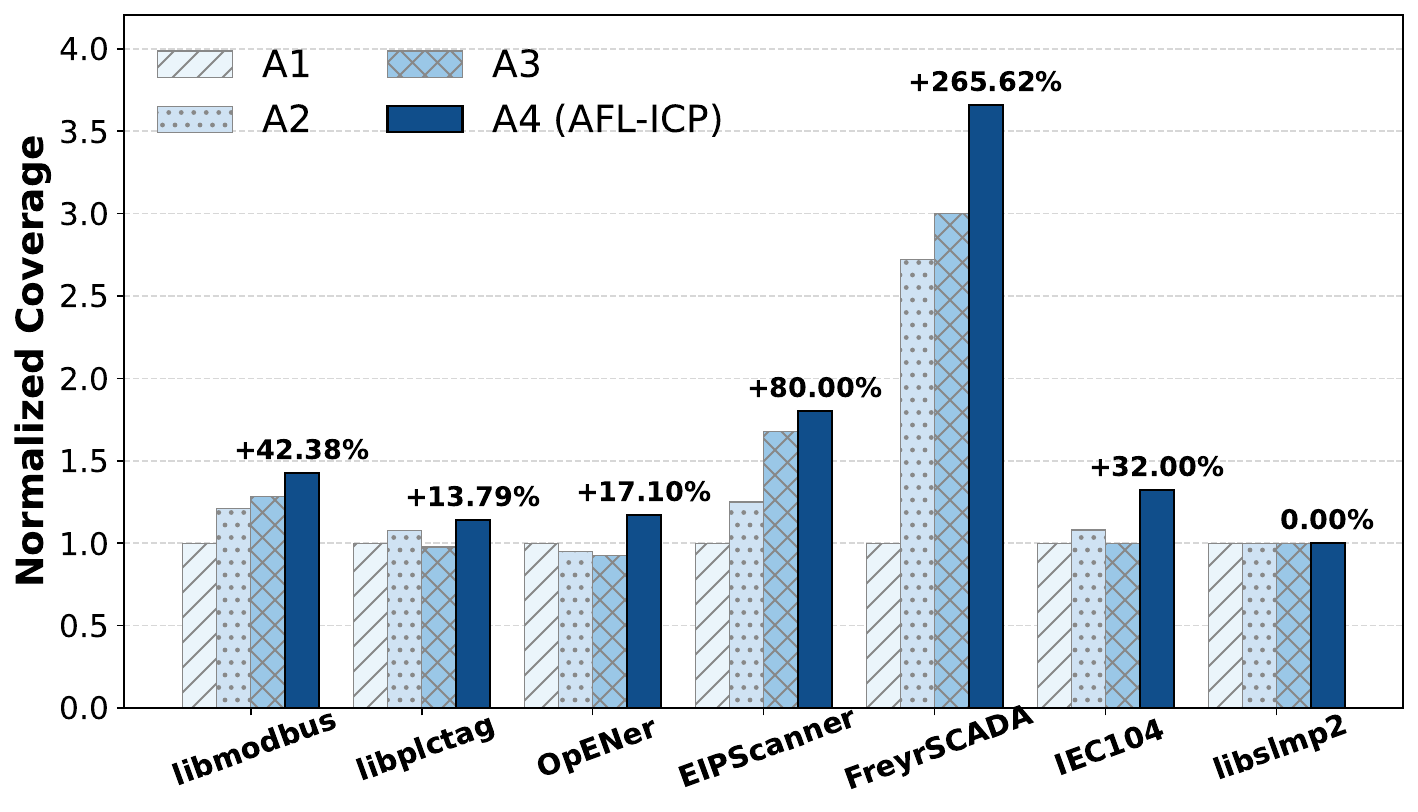}
    }
    \hfill
    \subfloat[Line Coverage\label{fig:ablation_line}]{
        \includegraphics[width=0.318\textwidth]{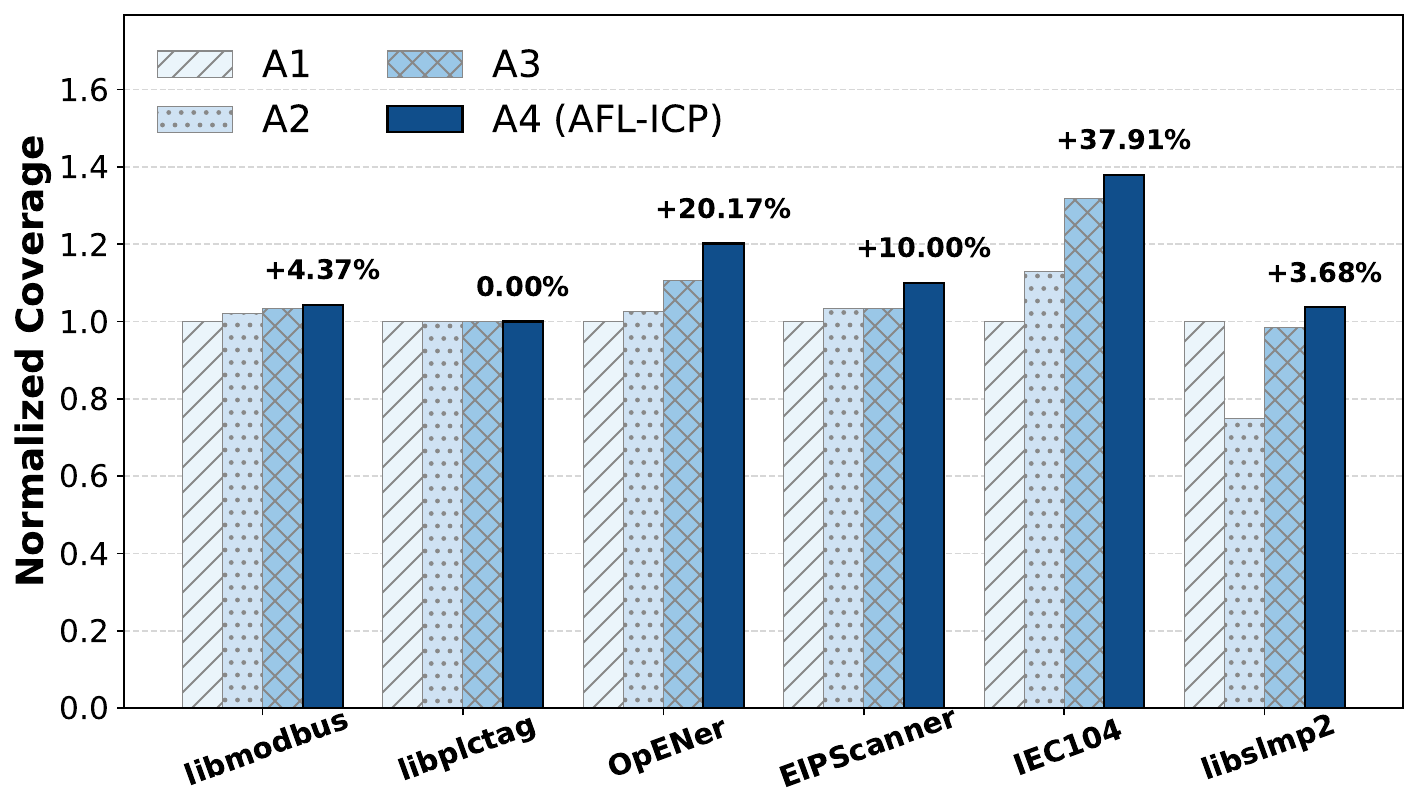}
    }
    \vspace{-0.1cm}
    \caption{Ablation study results showing state, transition, and line coverage improvements.}
    \label{fig:ablation_all}
    \vspace{-0.2cm}
\end{figure*}
\begin{figure}[htbp]
    \vspace{-0.3cm}
    \centering
    \includegraphics[width=0.35\textwidth]{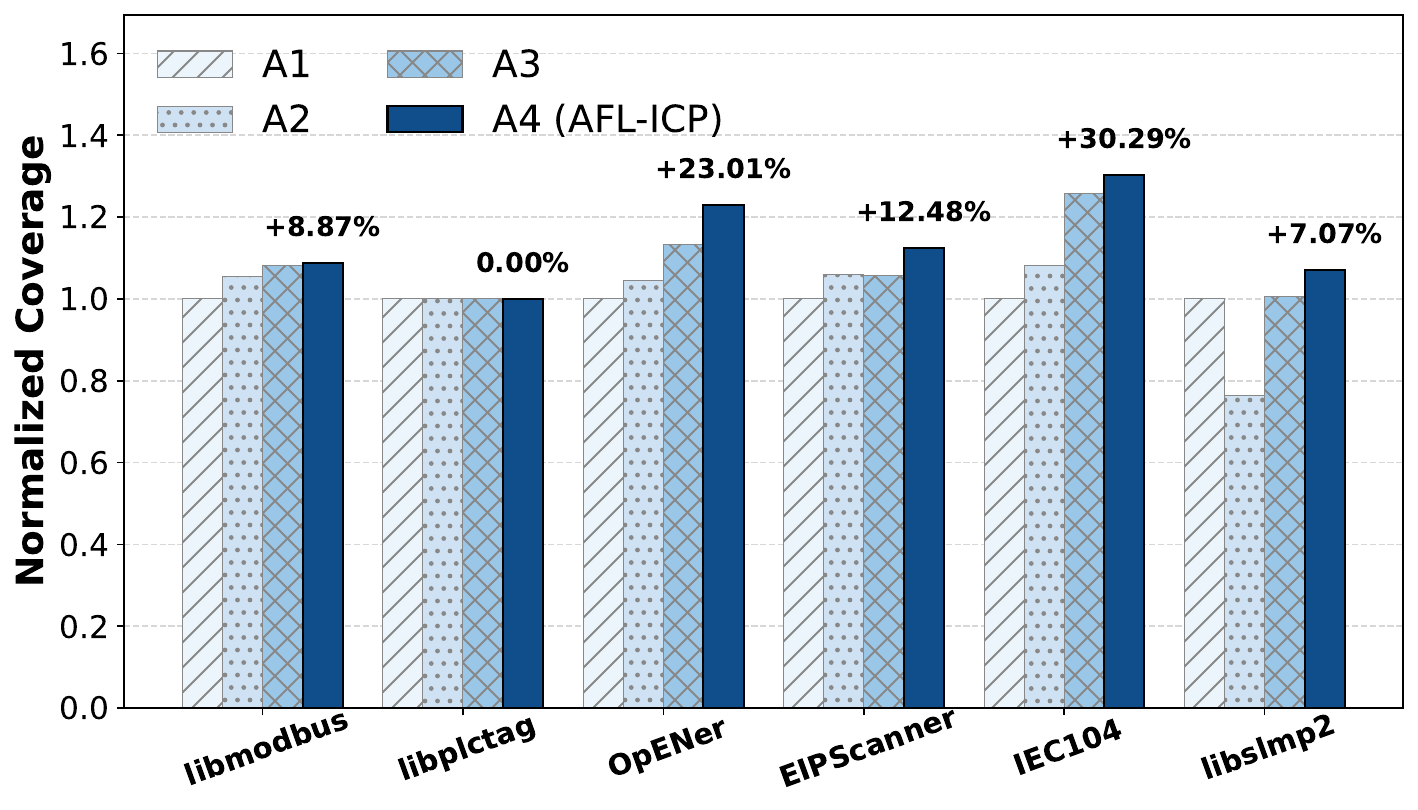}
    \vspace{-0.3cm}
    \caption{Ablation study results showing branch coverage improvements}
    \vspace{-0.6cm}
    \label{fig:ablation_branch}
\end{figure}
To evaluate the individual contributions of each component, we conducted an ablation study focusing on coverage improvements. 
Note that protocol adaptation provides the necessary execution environment, while semantic vulnerability reasoning focuses on vulnerability identification; neither directly influences coverage. 
Therefore, we specifically measure the coverage gains driven by \textit{specification formalization} and \textit{initial seed synthesis} using four incremental configurations:

\textbf{A1: AFLNet (Baseline)}. Standard AFLNet adapted for ICPs, without any LLM enhancements. This serves as our baseline.

\textbf{A2: A1 + LLM-based Seed Generation (No Specification)}. Adds LLM-based seed synthesis using only protocol names and pre-trained knowledge, testing baseline LLM capability. 

\textbf{A3: A1 + Simple Document Preprocessing + Initial Seed Synthesis}. LLMs generate seeds guided by specifications processed through basic PDF-to-text conversion (PyPDF2). This evaluates the impact of raw text lacking structural context. 

\textbf{A4: A1 + Specification Formalization + Initial Seed Synthesis (Full \fuzzername{})}. Integrates complete components. The structured ICP knowledge guides the LLM to generate high-quality seeds, demonstrating the full potential of specification-guided fuzzing. 


This design allows us to evaluate: (1) the contribution of LLM-based seed generation alone (A2 vs. A1), (2) the impact of basic document guidance (A3 vs. A2), (3) the added value of structured specification parsing over simple document conversion (A4 vs. A3), and (4) the overall improvement of the complete system (A4 vs. A1).
Fig. \ref{fig:ablation_all} and Fig. \ref{fig:ablation_branch} jointly summarize the results of the ablation study. 
In these figures, we present the normalized coverage, where the A1 (Baseline) is set to 1.0. 

\textbf{Analysis of State-Space Exploration.} 
Fig. \ref{fig:ablation_state} and Fig. \ref{fig:ablation_trans} show that while the full configuration (A4) consistently excels, ablated configurations (A2, A3) fluctuate, particularly in OpENer and IEC104. 
This instability stems from \textit{imprecise guidance}: A2 suffers from imprecise guidance driven by general LLMs, while A3 struggles with conflicting constraints from unstructured text. 
Although minor variations reflect the stochastic nature of fuzzing, A4's robust performance confirms that specification guidance and structured parsing are needed to reliably navigate complex ICPs. 

\textbf{Analysis of Code Coverage.} 
Fig. \ref{fig:ablation_line} and Fig. \ref{fig:ablation_branch} further expose the counterproductive effects of unguided LLMs. 
Specifically, A2 causes a sharp regression in libslmp2, where hallucinations produce invalid seeds that fail basic parsing. 
Additionally, identical results for libplctag across all stages confirm the ``test harness saturation'' noted in Sec. \ref{code}. 
The progression from A3 to A4 demonstrates that only by accurately capturing critical protocol fields via our multimodal pipeline can the LLM generate high-quality seeds that penetrate deep functional logic.

\subsubsection{Memory Safety Bug Detection}\label{bug1}
Table~\ref{tab:bugNum} reports the unique crash-inducing sequences discovered across five campaigns. 
\fuzzername{} discovered 87 unique sequences, outperforming AFLNet (49, +77.6\%) and ChatAFL (38, +129.0\%). 

Detailed analysis confirmed eight distinct memory safety vulnerabilities, summarized in Table~\ref{tab:bugList}. 
In FreyrSCADA, we identified memory leaks.
Although its closed-source nature prevented precise root cause analysis, the bug has been officially acknowledged by the vendor. 
For EIPScanner, \fuzzername{} and AFLNet found all three heap buffer over-read bugs, while ChatAFL found only one. 
Notably, in IEC104, \fuzzername{} uniquely discovered three null pointer dereference and DoS vulnerabilities, compared to only one by baselines.  
Furthermore, \fuzzername{} exclusively triggered a heap buffer over-read in libslmp2 (ID 8), which remained undetected by others.

These findings validate that \fuzzername{}'s specification-guided strategy enables deeper state exploration, uncovering vulnerabilities inaccessible to existing fuzzers. 

\subsubsection{Semantic and Logic Bug Detection}\label{bug2}
We evaluate \fuzzername{}'s capability to detect semantic deviations, categorized into three types:
(1) \textbf{Strict Non-Conformance (SNC)}, where the implementation explicitly contradicts protocol rules;
(2) \textbf{Fragile Error Handling (FEH)}, where the implementation adopts unsafe practices in ambiguous specification definition scenarios;
(3) \textbf{Implementation Logic Flaws (ILF)}, where fundamental code logic errors cause functional deviations. 
Through our analysis, we uncovered multiple vulnerabilities (Table \ref{tab:vul2}). 
To rule out LLM hallucinations, each flagged finding was validated by (1) replaying the request sequence to reproduce the deviant behavior on the target implementation, (2) locating the specific clause in the original protocol specification that the behavior violates, and (3) reporting the issue to upstream maintainers where applicable.

\textbf{SNC}. Violations of explicit protocol rules often undermine security boundaries. For instance, in libmodbus (Bug 1), the server processes packets with invalid Protocol IDs instead of discarding them. This violation of the ``filter-at-header'' principle allows malicious packets to potentially bypass firewalls or DPI systems. Similarly, in OpENer (Bug 8), the server erroneously replies to requests with non-zero status fields, exposing a side-channel for attackers to enumerate valid session handles.

\begin{table}[t!]
    \centering
    \vspace{-0.5cm}
    \caption{Unique crash-inducing sequences discovered.}
    \vspace{-0.3cm}
    \small
    \resizebox{\columnwidth}{!}{
    \setlength{\tabcolsep}{3.5pt}
    \begin{tabular}{l|c|cc|cc}
        \toprule
        \textbf{Program} & \textbf{\fuzzername{}} &\textbf{AFLNet}
        &\textbf{Improve} &\textbf{ChatAFL} &\textbf{Improve} \\
        \hline
        \hline
        libmodbus &0  &0  &0.00\%   &0  &0.00\% \\
        libplctag &0  &0  &0.00\%   &0  &0.00\% \\
        OpENer &0  &0  &0.00\%   &0  &0.00\% \\
        EIPScanner &49  &25 &96.00\%  &10  &390.00\%   \\
        FreyrSCADA &30  &21  &42.86\%  &27  &11.11\%   \\
        IEC104 &7  &3  &133.33\%  &1  &600.00\%   \\ 
        libslmp2 &1  &0  &-   &0  &- \\
        \hline
        \hline
        Total &87  &49  &77.55\%  &38  &128.95\% \\
        \bottomrule
    \end{tabular}
    }
    \vspace{-0.5cm}
    \label{tab:bugNum}
\end{table}

\textbf{FEH}. Unsafe practices in edge cases can lead to resource exhaustion. A critical example is found (Bug 4), where the server lacks application-layer timeouts when reading the header. Attackers can exploit this by establishing numerous connections that send incomplete headers, launching a Slowloris-style DoS attack that indefinitely occupies server threads. 

\textbf{ILF}. Fundamental logic errors often result in functional deviations. In libslmp2 (Bug 14), the implementation incorrectly treats stream-oriented TCP as message-oriented. It parses only the first frame in a received buffer and discards the rest, causing data loss when TCP stickiness occurs. Additionally, in libmodbus (Bug 2), implicit length reliance causes desynchronization, allowing attackers to inject ``ghost commands'' hidden within residual TCP data. 

These discoveries highlight that subtle semantic deviations, often overlooked by traditional crash-based fuzzers, can be effectively identified through \fuzzername{}.

\begin{table*}[thbp]
    \centering
    \caption{Summary of memory safety vulnerabilities discovered by \fuzzername{}}
    \vspace{-0.2cm}
    \resizebox{\textwidth}{!}{
    \begin{tabular}{c |c c |l}
        \hline
        \textbf{ID} & \textbf{Subject} & \textbf{Version} & \multicolumn{1}{c}{\textbf{Memory Safety Bug Description}} \\
        \hline
        \hline
        1 & FreyrSCADA &V21.06.008-89-g917706d &Memory leaks after memory copying \\
        2 & IEC104 &be6d841 &Firmware Backoff: Null pointer dereference (Iec104.c: 1214)  \\
        3 & IEC104 &be6d841 &Firmware Update: Null pointer dereference (Iec104.c: 1129)  \\
        4 & IEC104 &be6d841 &Denial of service due to the data finish error (clock\_nanosleep.c: 78)  \\
        5 & EIPScanner &1.3.0-33-g12c89a5 &Heap buffer over-read via unchecked vector size in bulk data copy (Buffer.cpp: 146)  \\ 
        6 & EIPScanner &1.3.0-33-g12c89a5 &Heap buffer over-read when reading low byte of uint16\_t (Buffer.cpp: 51)  \\
        7 & EIPScanner &1.3.0-33-g12c89a5 &Heap buffer over-read when reading high byte of uint16\_t (Buffer.cpp: 52)  \\ 
        8 & libslmp2 &v1.0.0 &Heap buffer over-read via unchecked Number of loopback data field (svrskel.c: 76)  \\
        \hline
    \end{tabular}
    }
    \label{tab:bugList}
\end{table*}

\begin{table*}[thbp]
    \centering
    \vspace{-0.2cm}
    \caption{Summary of semantic and logic vulnerabilities discovered by \fuzzername{}. Categories: Strict Non-Conformance (SNC), Fragile Error Handling (FEH), and Implementation Logic Flaws (ILF).}
    \vspace{-0.3cm}
    \small
    \resizebox{\textwidth}{!}{
    \setlength{\tabcolsep}{4pt}
    \begin{tabular}{c |c c |c |p{12cm}}
        \hline
        \textbf{ID} & \textbf{Subject} & \textbf{Version} & \textbf{Cat.} & \multicolumn{1}{c}{\textbf{Semantic and Logic Bug Description}} \\
        \hline
        \hline
        1 & libmodbus &v3.1.10-6-g5c14f13 &SNC &Packets with invalid Protocol Identifier not discarded\\ 
        2 & libmodbus &v3.1.10-6-g5c14f13 &SNC &Packet boundary identified by implicit function code byte count instead of explicit length field \\ 
        3 & libmodbus &v3.1.10-6-g5c14f13 &FEH &Blocks and disconnects on payload mismatch instead of sending Exception code. \\ 
        4 & libplctag &v2.6.12-0-gdeccfa3 &FEH &Server blocks indefinitely on incomplete MBAP headers, causing resource exhaustion.\\ 
        5 & IEC104 &be6d841 &SNC &Violation of state machine logic,  accepts I-frames without STARTDT activation \\
        6 & IEC104 &be6d841 &SNC &Failure to validate APDU length before processing leads to out-of-bounds memory access \\ %
        7 & IEC104 &be6d841 &SNC &Incomplete protocol implementation fails to support fundamental monitoring ASDUs \\
        8 & OpENer &v2.3-573-g4aa166a &SNC &Replies to requests containing a non-zero status field, violating the no-reply requirement \\
        9 & EIPScanner &1.3.0-33-g12c89a5 &FEH &Responds to truncated packets with fabricated fields due to missing header length validation.  \\
        10& EIPScanner &1.3.0-33-g12c89a5 &ILF &SendRRData packet construction errors and failure to validate malformed request packets \\
        11& EIPScanner &1.3.0-33-g12c89a5 &SNC &Replies to invalid UnRegisterSession packets, violating the no-reply requirement  \\
        12& EIPScanner &1.3.0-33-g12c89a5 &SNC &Returns SUCCESS for requests with non-zero Options field instead of discarding. \\
        13& libslmp2 &v1.0.0 &SNC &Silently discards unsupported commands without error information in response. \\
        14& libslmp2 &v1.0.0 &SNC &Response header contains mismatching Serial No. \\
        15& libslmp2 &v1.0.0 &ILF &Only parses the first frame in a received TCP buffer and silently discards the remaining data. \\
        16& libslmp2 &v1.0.0 &FEH &Returns success for Loopback Test despite mismatch between data count and actual payload.\\
        \hline
    \end{tabular}
    }
    \vspace{-0.5cm}
    \label{tab:vul2}
\end{table*}

\section{Related Work}
\textbf{Fuzzing for ICPs.} 
Traditional ICP fuzzers, such as Polar~\cite{luo2019polar} and other specialized ICP fuzzers~\cite{luo2020ics, zuo2022vulnerability}, rely on function-code awareness or traffic-based state machine inference. 
However, these methods struggle to explore deep logic paths that are defined in specifications but absent from the initial seed traffic.
In contrast, \fuzzername{} overcomes the semantic blindness by directly leveraging authoritative knowledge extracted from protocol specifications.

\textbf{Knowledge-Driven Fuzzing.}
Existing works attempt to incorporate knowledge through manual formalization or automated inference.
For instance, Sun et al.~\cite{sun2022ipspex} and TCP-Fuzz~\cite{zou2021tcp} extract semantics from network traffic, inherently limiting their scope to observable behaviors. While ProphetFuzz~\cite{wang2024prophetfuzz}  employs simplified parsing unsuitable for the complexity of multi-modal ICP specifications. \fuzzername{} bridges this gap by automating the transformation of unstructured, multi-modal documentation into executable models, effectively eliminating the formalization bottleneck.

\textbf{Specification-Centric Security Analysis.}
A complementary line of work analyzes protocol specifications themselves rather than implementations. CellularLint~\cite{rahman2024cellularlint}, for example, applies NLP techniques to detect internal inconsistencies within 4G/5G cellular standards. Such efforts target the specification-vs-specification axis and produce findings about the document, whereas \fuzzername{} targets the specification-vs-implementation axis and operationalizes specifications as the ground truth that drives end-to-end fuzzing and conformance checking of ICP implementations.

\textbf{LLM-Guided Protocol Fuzzing.} 
Recent research utilizes LLMs for seed generation~\cite{meng2024large}, mutation~\cite{xia2024fuzz4all}, and complex input synthesis~\cite{yang2025kernelgpt}. 
However, tools like ChatAFL~\cite{meng2024large} primarily use LLMs as auxiliary mutation engines, often lacking the precision required for specialized ICS protocols.
\fuzzername{} advances this paradigm by integrating LLMs as a core ``architect'' throughout the entire fuzzing lifecycle, ensuring both semantic correctness and deep state exploration.

\section{Conclusion}
In this paper, we presented \fuzzername{}, an AI-native fuzzing framework that bridges the critical gap between abstract protocol specifications and concrete implementations in ICPs.
By automating the transformation of unstructured, multimodal specifications into a rigorous unified protocol schema, \fuzzername{} overcomes the specification-blindness and the manual formalization bottleneck. 
Our approach systematically integrates LLMs into every phase of the workflow, effectively transforming fuzzing from a blind stochastic search into a specification-guided process.
Extensive evaluations demonstrate that \fuzzername{} significantly outperforms state-of-the-art fuzzers in coverage and uncovers 24 previously unknown vulnerabilities, including 16 semantic and logic bugs. 


\bibliographystyle{IEEEtran}
\bibliography{reference.bib}


\end{document}